\documentclass[twocolumn]{aastex631}

\accepted{January 18, 2023}
\submitjournal{ApJL}

\graphicspath{{./}{figures/}}

\begin{document}

\title{COSMOS2020: Discovery of a protocluster of massive quiescent galaxies at $z=2.77$}

\correspondingauthor{Kei~Ito}
\author[0000-0002-9453-0381]{Kei~Ito}
\altaffiliation{JSPS Research Fellow (PD)}
\email{kei.ito@astron.s.u-tokyo.ac.jp, kei.ito.astro@gmail.com}
\affiliation{Department of Astronomy, School of Science, The University of Tokyo, 7-3-1, Hongo, Bunkyo-ku, Tokyo, 113-0033, Japan}
\author{Masayuki~Tanaka}
\affiliation{National Astronomical Observatory of Japan, 2-21-1 Osawa, Mitaka, Tokyo, 181-8588, Japan}
\affiliation{Department of Astronomical Science, The Graduate University for Advanced Studies, SOKENDAI, 2-21-1 Osawa, Mitaka, Tokyo, 181-8588, Japan}
\author[0000-0001-6477-4011]{Francesco~Valentino}
\affiliation{Cosmic Dawn Center (DAWN), Denmark}
\affiliation{Niels Bohr Institute, University of Copenhagen, Jagtvej 128, DK-2200 Copenhagen N, Denmark}
\author{Sune~Toft}
\affiliation{Cosmic Dawn Center (DAWN), Denmark}
\affiliation{Niels Bohr Institute, University of Copenhagen, Jagtvej 128, DK-2200 Copenhagen N, Denmark}
\author{Gabriel~Brammer}
\affiliation{Cosmic Dawn Center (DAWN), Denmark}
\affiliation{Niels Bohr Institute, University of Copenhagen, Jagtvej 128, DK-2200 Copenhagen N, Denmark}
\author{Katriona~M.~L.~Gould}
\affiliation{Cosmic Dawn Center (DAWN), Denmark}
\affiliation{Niels Bohr Institute, University of Copenhagen, Jagtvej 128, DK-2200 Copenhagen N, Denmark}
\author{Olivier~Ilbert}
\affiliation{Aix Marseille Univ, CNRS, CNES, LAM, Marseille, France}
\author{Nobunari~Kashikawa}
\affiliation{Department of Astronomy, School of Science, The University of Tokyo, 7-3-1, Hongo, Bunkyo-ku, Tokyo, 113-0033, Japan}
\author{Mariko~Kubo}
\affiliation{Astronomical Institute, Tohoku University, Aoba-ku, Sendai 980-8578, Japan}
\author[0000-0002-2725-302X]{Yongming~Liang}
\affiliation{Institute for Cosmic Ray Research, The University of Tokyo, 5-1-5 Kashiwanoha, Kashiwa, Chiba 277-8582, Japan}
\author{Henry~J.~McCracken}
\affiliation{Institut d’Astrophysique de Paris, 98bis Boulevard Arago, F-75014, Paris, France}
\affiliation{Sorbonne Universit\'{e}s, UPMC Univ Paris 6 et CNRS, UMR 7095, Institut d’Astrophysique de Paris, 98 bis bd Arago, 75014 Paris, France}
\author[0000-0003-1614-196X]{John~R.~Weaver}
\affil{Department of Astronomy, University of Massachusetts, Amherst, MA 01003, USA}

\begin{abstract}
Protoclusters of galaxies have been found in the last quarter century. However, most of them have been found through the overdensity of star-forming galaxies, and there had been no known structures identified by multiple spectroscopically confirmed quiescent galaxies at $z>2.5$. In this letter, we report the discovery of an overdense structure of massive quiescent galaxies with the spectroscopic redshift $z=2.77$ in the COSMOS field, QO-1000. We first photometrically identify this structure as a $4.2\sigma$ overdensity with 14 quiescent galaxies in $7\times4\ {\rm pMpc^2}$ from the COSMOS2020 catalog. We then securely confirm the spectroscopic redshifts of 4 quiescent galaxies by detecting multiple Balmer absorption lines with Keck/MOSFIRE. All the spectroscopically confirmed members are massive ($\log{(M_\star/M_\odot)}>11.0$) and located in a narrow redshift range ($2.76<z<2.79$). Moreover, three of them are in the $1\times1\ {\rm pMpc^2}$ in the transverse direction at the same redshift ($z=2.760-2.763$). Such a concentration of four spectroscopically confirmed quiescent galaxies implies that QO-1000 is $>68$ times denser than in the general field. In addition, we confirm that they form a red sequence in the $J-K_s$ color. This structure's halo mass is estimated as $\log{(M_{\rm halo}/M_\odot)}>13.2$ from their stellar mass. Similar structures found in the IllustrisTNG simulation are expected to evolve into massive galaxy clusters with $\log{(M_{\rm halo}/M_\odot)}\geq14.8$ at $z=0$. These results suggest that QO-1000 is a more mature protocluster than the other known protoclusters. It is likely in a transition phase between the star-forming protoclusters and the quenched galaxy clusters.
\end{abstract}


\section{Introduction} \label{sec:intro}
\par In the local universe, it is widely known that the properties of galaxies are largely affected by their surrounding environments. In particular, massive and quiescent galaxies often reside in massive and dense structures, such as galaxy cluster cores \citep[e.g.,][]{Peng2010}. Exploring massive structures in the high redshift universe can unveil their origin. In recent decades, protoclusters, which are progenitors of local clusters with halo mass of $\log{(M_{\rm halo}/M_\odot)>14}$, have been found to $z\sim7-8$ \citep[e.g.,][]{Harikane2019, Hu2021, Laporte2022}. They have often been found through the overdensity of star-forming galaxies, such as bright galaxies in rest-frame ultraviolet continuum \citep[e.g.,][]{Steidel1998, Overzier2008, Toshikawa2018}, Ly$\alpha$ emission \citep[e.g.,][]{Shimasaku2003, Jiang2018, Harikane2019}, H$\alpha$ emission \citep[e.g.,][]{Hayashi2012, Darvish2020}, and  infrared \citep[e.g.,][]{Oteo2018, Miller2018}.
\par In the general field, recent multi-wavelength surveys and near-infrared spectrographs have identified galaxies with suppressed star formation activities even at high redshifts \citep[e.g.,][]{Kriek2009, Glazebrook2017, Forrest2020a, Valentino2020, DEugenio2021, Marchesini2022}. These quiescent galaxies have been spectroscopically confirmed up to $z=4.01$ \citep[][]{Tanaka2019}, and the James Webb Space Telescope has discovered candidates even at $z>4$ \citep[][]{Carnall2022}. They are thought to have intense star formation in the first $1-2$ Gyrs from the Big Bang and suddenly quenched \citep[][]{Schreiber2018, Belli2019, Forrest2020a, Forrest2020b, Saracco2020, Valentino2020}. However, the relationship between their suppressed star formation activity and the surrounding environment is not understood yet. Some studies report an individual quiescent galaxy in protoclusters selected from star-forming galaxies \citep[][]{Kubo2021, Kalita2021} and the high quiescent fraction for a protocluster using both photo-$z$ and spec-$z$ galaxy samples \citep[][]{Spitler2012, McConachie2021}. However, most protoclusters have been found through the distribution of star-forming galaxies, and there had not been any protocluster at $z>2.5$ identified by the distribution of only massive quiescent galaxies, as is done in local clusters. This situation could induce sample bias in the dense structure sample and prevent us from understanding the relationship between quenching and the surrounding environment.
\par Finding overdense regions through the distribution of quiescent galaxies has been challenging because the number density of quiescent galaxies is small ($\sim10^{-5}{\rm Mpc^{-3}}$ at $z\geq3$) compared to that of star-forming galaxies \citep[e.g.,][]{Muzzin2013, Straatman2014, Davidzon2017, Schreiber2018, Merlin2019}, leading to less statistical significance in overdensity. The Cosmic Evolution Survey \citep[COSMOS,][]{Scoville2007} can solve this issue. Thanks to the deep multi-band photometry across the $\sim2\ {\rm deg^{2}}$ from the latest COSMOS2020 catalog \citep{Weaver2022}, we can select quiescent galaxies even in the low mass regime ($\log{(M_\star/M_\odot)}\sim10-10.5$) with the precise photometric redshift from the wide field.
\par In this letter, we report the discovery of an overdense structure of massive quiescent galaxies at $z=2.77$ with four spectroscopically confirmed quiescent galaxies from Keck/MOSFIRE observations in the COSMOS field. This structure is a good example of a mature protocluster already dominated by quiescent galaxies at less than 2.5 Gyr after the Big Bang. This letter is organized as follows. We first summarize our target selection for overdense regions of quiescent galaxies and our Keck/MOSFIRE observations in Section \ref{sec:2}. Then, the analysis of the observed data and the discussion based on its result are summarized in Section \ref{sec:3}. Lastly, we conclude this letter in Section \ref{sec:4}. We assume the following cosmological parameters: $H_0 = 70\ {\rm km\ s^{-1}\ Mpc^{-1}}$, $\Omega_m = 0.3$, and $\Omega_\Lambda = 0.7$. Magnitudes are in the AB system \citep{Oke1983}. We assume the initial mass function of \citet{Chabrier2003}.
\section{Data} \label{sec:2}
\subsection{Selection of quiescent galaxy overdensity} \label{sec:2-1}
\par We search overdense structures of quiescent galaxies at $z\sim3$ in the COSMOS field in $\sim2\ {\rm deg^2}$ based on the projected distribution of quiescent galaxies. The galaxy sample is based on the SED fitting with {\tt MIZUKI} \citep{Tanaka2015} to the photometry of the COSMOS2020 \textsc{Classic} catalog \citep{Weaver2022}, covering the wide wavelength with 40 bands from $u$ band to IRAC ch4. Three criteria are applied to select quiescent galaxies. First, a cut in UltraVISTA/$K_s$ 3\arcsec aperture magnitude is applied as $K_s<24.8$ mag to minimize the effect of the depth inhomogeneity when measuring the galaxy number density. This threshold corresponds to the $3\sigma$ limiting magnitude in the ``deep" stripe of UltraVISTA. Then, we select quiescent galaxies based on their $1\sigma$ upper limit of specific star formation rate (sSFR) from SED fitting ($\log{({\rm sSFR}_{\rm 1\sigma, upper}/{\rm yr^{-1}})}<-9.5$), following to our previous works \citep[e.g.,][]{Kubo2018, Tanaka2019, Valentino2020, Ito2022}. Galaxies not satisfying this criterion are classified as star-forming galaxies. This threshold corresponds to $\sim1$ dex lower than the sSFR of the star formation main sequence \citep[e.g.,][]{Tomczak2016, Leslie2020}. Lastly, we put the stellar mass cut as $\log{(M_\star/M_\odot)}>10.3$ since the magnitude cut leads to incompleteness in terms of stellar mass. This threshold corresponds to a 90\% completeness limit of quiescent galaxies at our target redshift ($z\sim2.8$), based on our $K_s$ magnitude limit and the method commonly used in previous works \citep[e.g.,][]{Pozzetti2010, Laigle2016, Weaver2022}.
\par The overdensity map of quiescent galaxies is obtained by the Gaussian kernel density estimation method \citep[e.g.,][]{Badescu2017, Chartab2020, Ito2021}. We construct a redshift slice with a width of $\delta z=\pm0.1$, comparable to the scatter of our photometric redshift of $z>2$ quiescent galaxies \citep[c.f.,][]{Ito2022}, and the galaxy distribution is smoothed with the bandwidth (standard deviation of Gaussian) as 3 arcmin  ($\sim5.5$ cMpc at $z\sim2.8$), which is equivalent to the typical half-mass radius of protoclusters at this redshift estimated in simulations \citep{Chiang2013}. Regions near bright stars are masked out based on the flag in the COSMOS2020 catalog ({\tt FLAG\_COMBINED}). In addition, the number density near the edges of the survey region and the masks are corrected, following the method in \citet{Chartab2020}. We ignore the area where the correction factor is more than 1.67. After careful screening of the overdense regions with the number of quiescent galaxies, we find multiple significant overdense structure candidates of quiescent galaxies at $2.5<z_{\rm phot}<3.5$, corresponding to the survey volume of $1.5\times10^7{\rm cMpc^{3}}$.
\par
\begin{figure*}
    \centering
    \includegraphics[width=17cm]{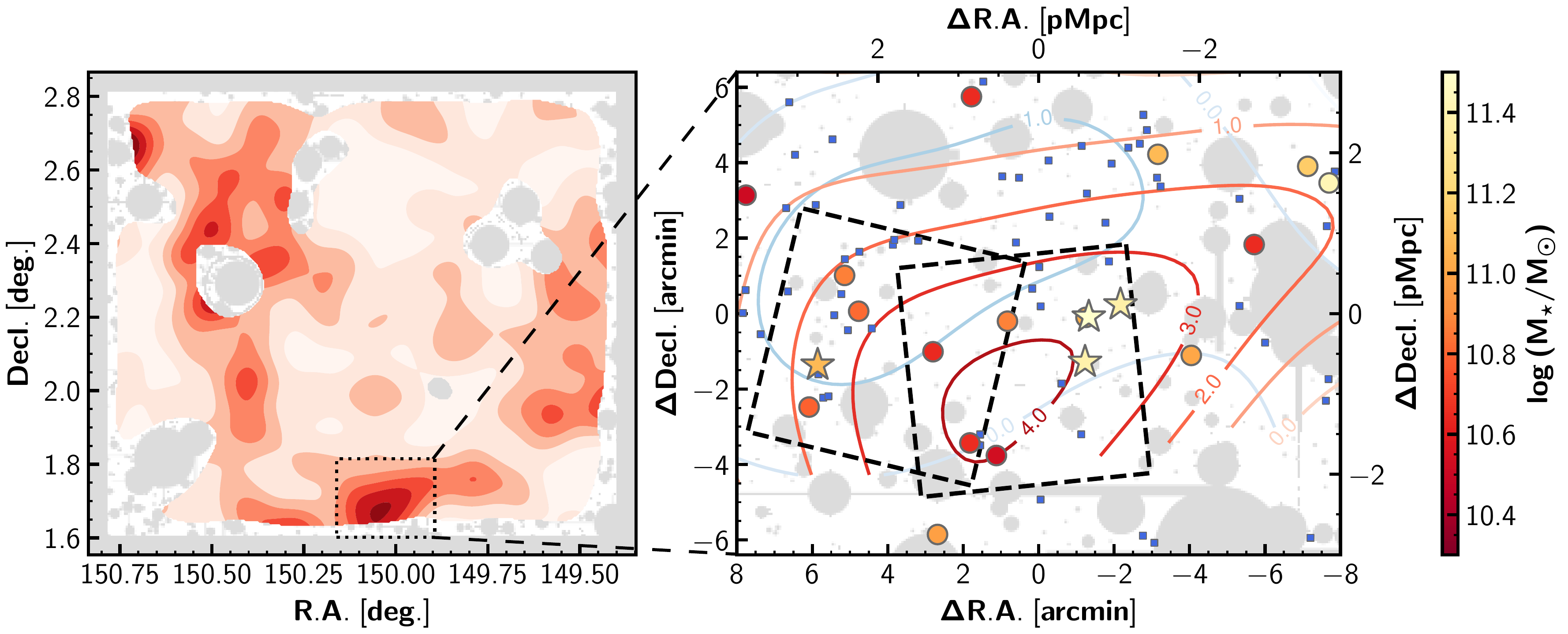}
    \caption{Left panel: Density map of quiescent galaxies with $\log{(M_\star/M_\odot)}>10.3$ at $2.74<z_{\rm phot}<2.94$. The red shaded contours indicate $-1\sigma\ ,0\sigma,\ 1\sigma,\ 2\sigma,\ 3\sigma,\ 4\sigma$ overdensity. Gray regions correspond to the mask or the outside of the survey area \citep[c.f.,][]{Weaver2022}. The dashed rectangle indicates the overdense structure reported in this paper. Right panel: Zoom in view of the overdense structure. The red and blue contours correspond to the density map of quiescent and star-forming galaxies with the same stellar mass cut, respectively. The number in these contours indicates the overdensity significance. The stars are the quiescent galaxies spectroscopically confirmed in this study, and the circles are photometrically selected ones at $2.74<z_{\rm phot}<2.94$. They are color-coded by their stellar mass. The blue squares are star-forming galaxies in the same redshift slice. Black squares show the location of the MOSFIRE masks.}
    \label{fig:1}
\end{figure*}

 A significant overdense region of quiescent galaxies is found in the redshift slice of $2.74<z_{\rm phot}<2.94$ (Figure \ref{fig:1}). According to the $2\sigma$ contour, this region extends $14\times 8\ {\rm arcmin^2}$ ($\sim 7\times4\ {\rm pMpc^2}$ at this redshift). Hereafter, we define this as the area of this overdense region. It has an overdensity, defined as the number density excess over its average normalized by the average, of $4.2\sigma$ at the peak. There are 14 quiescent galaxies, and they are all as massive as $\log{(M_\star/M_\odot)}>10.5$ (Table \ref{tab:1}).
\par A point that we should mention about this structure is that star-forming galaxies are not strongly concentrated. The overdensity is also measured from star-forming galaxies, which is $\sim 1\sigma$ at most (see right panel of Figure \ref{fig:1}). At the density peak of quiescent galaxies, it is $\sim0\sigma$. This implies that this structure could not be found based on the distribution of star-forming galaxies. The fraction of quiescent galaxies of this structure is $0.34\pm0.11$ at $\log{(M_\star/M_\odot)}>10.3$ in this redshift slice, which is $\sim3$ times higher than the average in the entire COSMOS field ($0.129\pm0.009$). This high value suggests that quenching is more efficient in this structure compared to the general field.
\subsection{Keck/MOSFIRE observation and data reduction}
\par We conduct Keck/MOSFIRE $H$-band spectroscopy targeting the quiescent galaxies inside this structure. Using two masks, we observe $9$ quiescent galaxies at $2.74<z_{\rm phot}<2.94$. The first mask (R.A., Decl. = 10:00:08.26, +01:40:58.64) was observed on the first half night of 19th April 2022, and the other (R.A., Decl. = 10:00:22.82  +01:41:36.54) was observed on the first half night of 20th April 2022. Each exposure was 120 s long, and ABBA nodding was applied. A bright star was included in each mask, and exposures, where the star had a significantly lower signal-to-noise ratio than the other masks, were not used in this study. The final total integration times for these masks were 3.38 hr and 2.65 hr, respectively. The obtained data was reduced by {\tt MOSFIRE DRP}\footnote{\url{https://keck-datareductionpipelines.github.io/MosfireDRP/}}, and the 1D spectra were optimally extracted from the reduced 2D spectra \citep[][]{Horne1986}. Spectra of A0V stars were used to convert the pixel count to the flux density. The flux loss due to the slit was corrected using the UltraVISTA/$H$ band magnitude in the COSMOS2020 catalog. The noise spectra were estimated as the standard deviation of the pixel count where no object is observed for each mask.
\par We attempt to measure the redshift from these 1D spectra by fitting the spectral template with {\tt SLINEFIT}\footnote{\url{https://github.com/cschreib/slinefit}} \citep{Schreiber2018}. In {\tt SLINEFIT}, the stellar continuum templates, identical to those used in {\tt EAZY} \citep[][]{Brammer2008}, are employed. These templates are convolved with a Gaussian velocity profile with the stellar velocity dispersion estimated from the spectra by {\tt ppxf} \citep{Cappellari2017}. The velocity dispersion of galaxies will be discussed in another paper in detail (Ito et al. in prep.). For galaxies without successful velocity dispersion measurement, we use the velocity dispersion value expected from their stellar mass according to the empirical relation in \citet{Schreiber2018}, which is based on the velocity dispersion and the stellar mass of \citet[][]{Belli2017}. We explore the redshift solution at $2<z<4$, which covers most of the redshift range suggested by the photometric SED fitting. In the fitting, pixels largely affected by the skylines are not used. If the spectra do not have enough signal-to-noise ratio, they are re-binned by $4\times$ to a resolution of 6.5\AA. The obtained redshifts do not change significantly, even if we use the native resolution. 
\section{Results and Discussion} \label{sec:3}
\subsection{Spectroscopic identification of four quiescent galaxies} \label{sec:3-1}
\par
\begin{figure*}
    \centering
    \includegraphics[width=17cm]{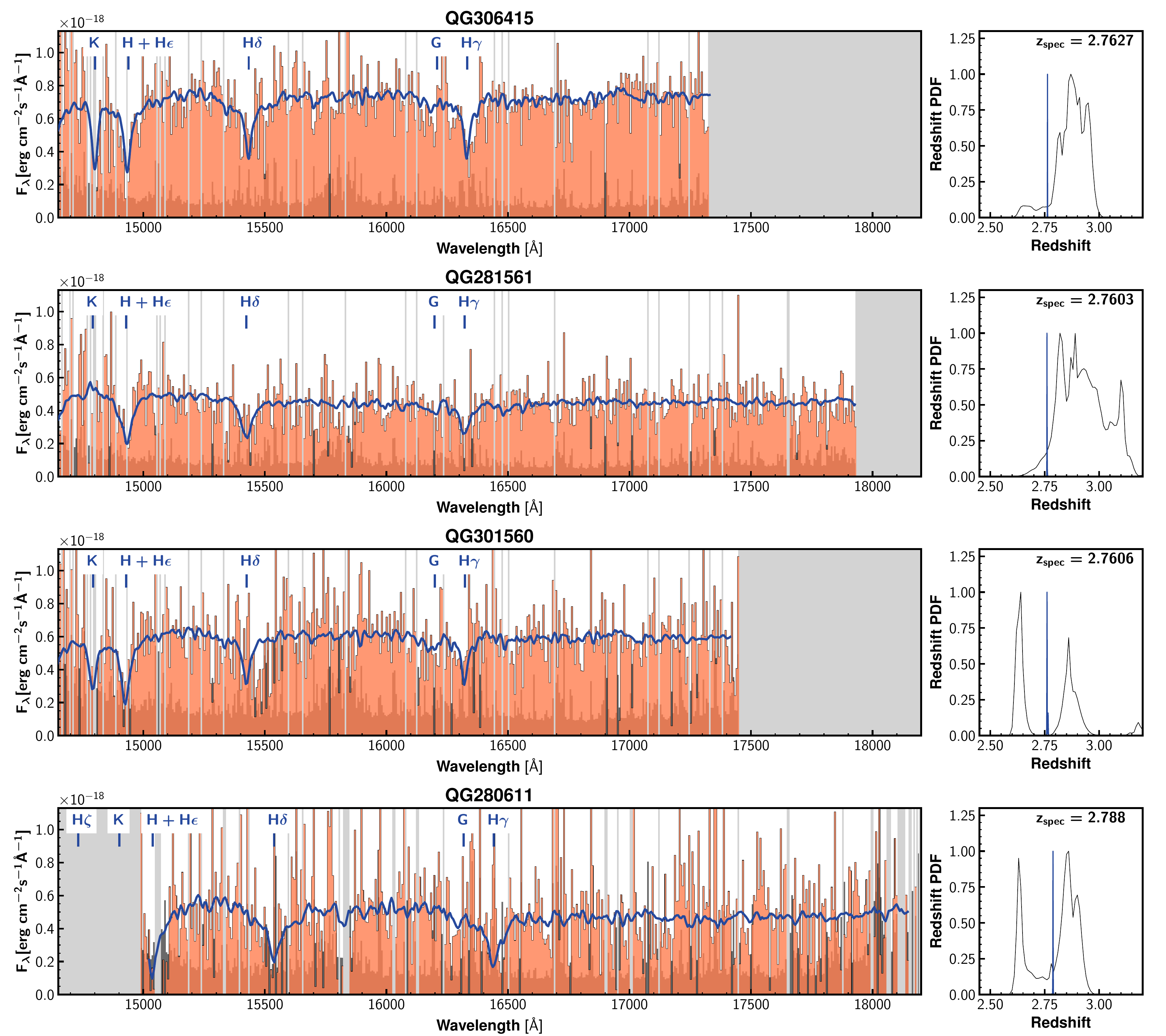}
    \caption{Left panel: Spectra of four quiescent galaxies with significant absorption lines. The object spectra are shown with orange shading, and the noise spectra are shown in dark shading on each panel. In addition to QG301560 and QG280611, whose spectral analyses were conducted with binned spectra, the others' spectra are also re-binned over 4 pixels just for illustrative purposes. The blue lines are the best fit from {\tt SLINEFIT}. The gray-masked regions are not observed or largely affected by the skylines and not used for the fitting. Right panel: Blue and black lines indicate the probability distribution of the spectroscopic redshift from {\tt SLINEFIT} and the photometric redshift from {\tt MIZUKI}, respectively.}
    \label{fig:2}
\end{figure*}

 Among $9$ spectra, we observe multiple significant absorption lines in four spectra (Figure \ref{fig:2}). The other galaxies were too faint to detect their absorption lines, and their redshifts were not constrained from their spectra. Two galaxies, QG306415 and QG281561, have enough signal-to-noise ratio even without re-binning (median ${\rm S/N}=4.6\ {\rm and}\  3.2$ per pixel, respectively), whereas the other two, QG301560 and QG280611, have enough signal-to-noise ratio after 4-pixel re-binning  (median ${\rm S/N}=3.3\ {\rm and}\ 2.6$ per pixel, respectively). 
\par Interestingly, all of their best redshifts are concentrated in the small redshift range ($z\sim2.76-2.79$, Table \ref{tab:1}). Following the method in \citet{Schreiber2018}, we check the probability that the true redshift lies within $\pm 0.01$ from the best-fit redshift by integrating the redshift probability distribution function. The value is high for all four galaxies (96.4-100\%), implying that these estimations are robust. Therefore, we conclude that this observation confirms that this structure, hereafter called QO-1000, has multiple quiescent galaxies in the small redshift range. We employ $z=2.77$ as the redshift of QO-1000, the mean of the redshift of these four. Hereafter, these four quiescent galaxies are called spectroscopic quiescent members, and the other ten galaxies inside QO-1000 are called photometric quiescent members. The latter members are also plausible since it is possible we could not detect their absorption lines just because of the depth limit of the observation.
\par There are three points that we should make for these four spectroscopic quiescent members. First, they are close to each other and located in $8.0\times1.6\ {\rm arcmin^2}$ ($\sim4\times1\ {\rm pMpc^2}$). In particular, three out of four (QG306415, QG281561, QG301560) are located in $1\times1\ {\rm pMpc^2}$ at the same redshift of $z=2.76$. Second, they are all massive. They have the stellar mass of $\log{(M_\star/M_\odot)}=11.08-11.53$, which is more massive than the peak stellar mass of the stellar mass function of quiescent galaxies at this redshift \citep[e.g.,][]{Davidzon2017}. Lastly, even if we conduct the SED fitting again with the spectroscopic redshift, they have a low specific star formation rate (see Table \ref{tab:1}), which satisfies the criteria of the quiescence (c.f., Section \ref{sec:2-1}) and is $>1$ dex lower than that of the star formation main sequence at this redshift \citep[e.g.,][]{Leslie2020}. One galaxy, QG281561, is observed in ALMA Band 7 but not detected \citep[A3COSMOS,][]{Liu2019}. Assuming the modified black body dust SED with the typical dust temperature of sub-millimeter galaxies \citep[][]{Dudzeviciute2020}, its $3\sigma$ upper limit suggests that the SFR of QG281561 is $\log({\rm SFR}/M_\odot\ {\rm yr}^{-1})<1.74$, supporting the value from the SED fitting.

\begin{deluxetable*}{clllllllll}
\tablecaption{Properties of quiescent galaxies in QO-1000}\label{tab:1}
\tablewidth{0pt}
\tablehead{
\colhead{ID\tablenotemark{a}} & \colhead{R.A. (J2000)} & \colhead{Decl. (J2000)} & \colhead{$z_{\rm spec}$\tablenotemark{b}} &\colhead{$p_z$\tablenotemark{c}} &\colhead{$z_{\rm phot}$}& \colhead{H\tablenotemark{d}}  & \colhead{$\log{M_\star}$ \tablenotemark{e}} & \colhead{$\log{\rm SFR_{\rm SED}}$\tablenotemark{e}}  &\colhead{$\log{\rm sSFR_{\rm SED}}$\tablenotemark{e}}  \\
\colhead{} & \colhead{[deg.]} & \colhead{[deg.]} &\colhead{} & \colhead{} & \colhead{} & \colhead{[mag]} & \colhead{[$\log{(M_\odot)}$]}&\colhead{[$\log{(M_\odot\ {\rm yr^{-1}})}$]}&\colhead{[$\log{({\rm yr^{-1}})}$]}}
\decimalcolnumbers
\startdata
QG306415 & 149.99164 & 1.71213  & $2.7627^{+0.0003}_{-0.0005}$ & 1.000 & $2.87^{+0.04}_{-0.03}$&21.92 & $11.361^{+0.008}_{-0.002}$ & $0.772^{+0.307}_{-0.005}$&$-10.59^{+0.31}_{-0.01}$\\
QG281561 & 150.00727 & 1.68677 & $2.7603^{+0.0008}_{-0.0003}$ & 1.000 & $2.92^{+0.05}_{-0.05}$&22.39 &$11.339^{+0.003}_{-0.004}$ & $1.548^{+0.006}_{-0.238}$&$-9.79^{+0.01}_{-0.24}$ \\
QG301560 & 150.00561 & 1.70671 & $2.7606^{+0.002}_{-0.0007}$ & 0.964 & $2.84^{+0.02}_{-0.21}$&22.22 & $11.532^{+0.003}_{-0.021}$ & $1.816^{+0.007}_{-0.187}$&$-9.72^{+0.03}_{-0.19}$\\
QG280611 & 150.12540 & 1.68547 & $2.7880^{+0.0006}_{-0.0007}$ & 0.998 & $2.80^{+0.06}_{-0.10}$&22.31 & $11.080^{+0.002}_{-0.002}$ & $0.490^{+0.004}_{-0.004}$&$-10.590^{+0.006}_{-0.006}$\\
\hline
QG300476 & 150.00687 & 1.70650 & - & - &$2.81^{+0.17}_{-0.22}$ & 22.92 & $11.04^{+0.07}_{-0.10}$ & $0.86^{+0.19}_{-0.16}$ & $-10.17^{+0.29}_{-0.23}$ \\
QG283763 & 149.96034 & 1.68968 & - & - &$2.93^{+0.17}_{-0.21}$ & 23.37 & $11.00^{+0.08}_{-0.08}$ & $0.15^{+0.30}_{-0.13}$ & $-10.86^{+0.38}_{-0.21}$ \\
QG298647 & 150.04154 & 1.70480 & - & -  &$2.80^{+0.19}_{-0.15}$ & 23.63 & $10.88^{+0.07}_{-0.07}$ & $-0.13^{+0.23}_{-0.30}$ & $-11.0^{+0.30}_{-0.37}$ \\
QG319735 & 150.11350 & 1.72504 & - & -  &$2.79^{+0.08}_{-0.06}$ & 22.93 & $10.87^{+0.02}_{-0.02}$ & $0.91^{+0.06}_{-0.15}$ & $-9.95^{+0.08}_{-0.18}$ \\
QG303434 & 150.10729 & 1.70921 & - & - &$2.93^{+0.08}_{-0.08}$ & 23.26 & $10.82^{+0.04}_{-0.03}$ & $0.55^{+0.25}_{-0.05}$ & $-10.27^{+0.28}_{-0.08}$ \\
QG261850 & 150.12919 & 1.66686 & - & - &$2.86^{+0.10}_{-0.11}$ & 23.29 & $10.80^{+0.06}_{-0.04}$ & $0.93^{+0.11}_{-0.17}$ & $-9.87^{+0.15}_{-0.23}$ \\
QG246253 & 150.05815 & 1.65102 & - & - &$2.81^{+0.18}_{-0.17}$ & 23.52 & $10.65^{+0.08}_{-0.08}$ & $0.35^{+0.22}_{-0.22}$ & $-10.31^{+0.30}_{-0.30}$ \\
QG332672 & 149.93256 & 1.73868 & - & - &$2.81^{+0.21}_{-0.3}$ & 25.45 & $10.65^{+0.09}_{-0.12}$ & $0.37^{+0.24}_{-0.18}$ & $-10.28^{+0.36}_{-0.27}$ \\
QG285514 & 150.07444 & 1.69131 & - & - &$2.78^{+0.16}_{-0.19}$ & 23.55 & $10.64^{+0.07}_{-0.1}$ & $0.80^{+0.17}_{-0.17}$ & $-9.85^{+0.27}_{-0.25}$ \\
QG241388 & 150.04645 & 1.64546 & - & - &$2.77^{+0.22}_{-0.22}$ & 23.94 & $10.53^{+0.11}_{-0.11}$ & $0.27^{+0.22}_{-0.19}$ & $-10.27^{+0.33}_{-0.30}$ \\
\enddata
\tablenotetext{a}{The object IDs are taken from the COSMOS2020 \textsc{Classic} catalog \citep[][]{Weaver2022}. Spectroscopically confirmed ones are sorted by the signal-to-noise ratio of the spectra, whereas the others are sorted by their stellar mass.}
\tablenotetext{b}{Spectroscopic redshift from this MOSFIRE observation}
\tablenotetext{c}{The interpreted probability within $\pm 0.01$ from the best-fit spectroscopic redshift}
\tablenotetext{d}{The $H$ band Kron magnitude from COSMOS2020 \textsc{Classic} catalog \citep{Weaver2022}}
\tablenotetext{e}{If the spectroscopic redshift is provided, these parameters are inferred with the redshifts fixed to $z_{\rm spec}$.}
\end{deluxetable*}
\subsection{Overdensity of quiescent galaxies}\label{sec:3-2}
\par In this subsection, we discuss how concentrated quiescent galaxies are in QO-1000 compared to the general field using spectroscopic redshifts. As reported in the previous section, QO-1000 has four quiescent galaxies concentrated in $\sim4\times1\ {\rm pMpc^2}$ in the transverse direction and $dz\sim0.03$ (corresponding to $2400$ km/s) in the line of sight direction. Assumed a cube-like shape, this structure has the volume of $1.1\times10^3\ {\rm cMpc^3}$, meaning that the number density of quiescent galaxies in QO-1000 is $3.4\times10^{-3} {\rm cMpc^{-3}}$.
\par In the same definition of quiescent galaxies with the stellar mass cut of $\log{(M_\star/M_\odot)}>11.0$, there are 360 quiescent galaxies at $2.5<z_{\rm phot}<3.0$ in the entire COSMOS field, giving the average number density of quiescent galaxies in the COSMOS field as $(4.7\pm0.3)\times10^{-5} {\rm cMpc^{-3}}$. This value implies that QO-1000 is $72.1\pm3.8$ times denser in quiescent galaxies than the average.
\par This high overdensity suggests that QO-1000 is a very dense structure of quiescent galaxies, which is supported by the precise spectroscopic redshifts. We note that this is based only on the spectroscopically confirmed galaxies. Since there are other photometric quiescent galaxies likely located in QO-1000, QO-1000 can be denser than this derived value. Therefore, taking the lower limit, we argue that this structure is at least 68 times denser than the general field.
\subsection{Color-magnitude diagram}
\par
\begin{figure*}
    \centering
    \includegraphics[width=16cm]{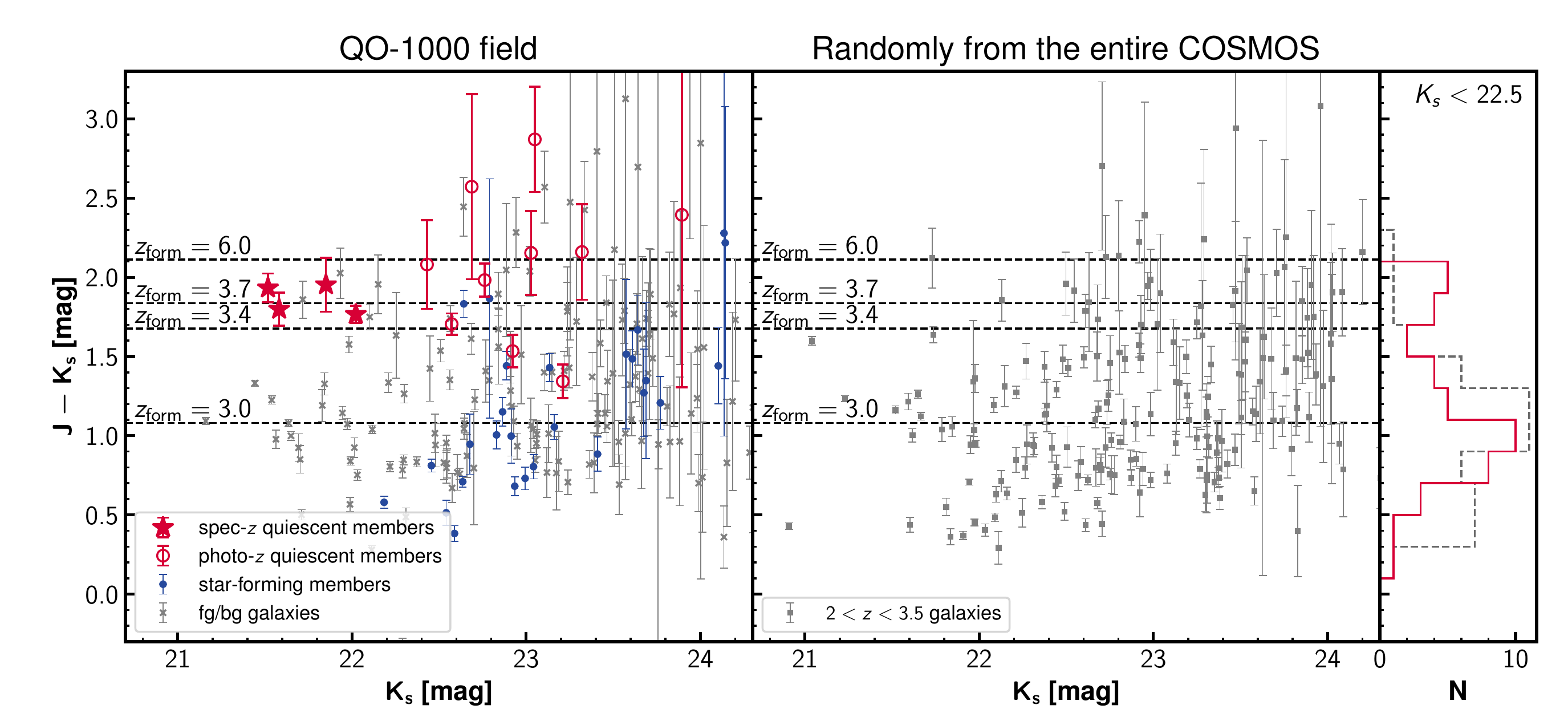}
    \caption{Left panel: Color-magnitude diagram for galaxies in QO-1000 field. The vertical axis shows $J-K_s$, and the horizontal axis shows $K_s$ band magnitude. The red stars and open circles represent spectroscopic and photometric quiescent members of QO-1000, respectively. Blue circles are star-forming members. The gray crosses represent the other galaxies at $2.0<z_{\rm phot}<3.5$ within the $2\sigma$ contour of Figure \ref{fig:1} in the projected space. The horizontal dashed lines show the $J-K_s$ color of the single-burst model of \citet{Bruzual2003} with different formation redshift ($z_{\rm form}=3.0,\ 3.4,\ 3.7,\ {\rm and}\ 6$). Central panel: Same as the left panel, but for galaxies at $2<z_{\rm phot}<3.5$ randomly drawn from the entire COSMOS field except QO-1000. The same number of objects as the QO-1000 field is drawn. Right panel: Histogram of $J-K_s$ of all galaxies with $K_s<22.5$ mag in QO-1000 field (red solid line) and randomly in the entire COSMOS (gray dashed line).}
    \label{fig:color-mag}
\end{figure*}

 Galaxy clusters at $z\leq2$ often exhibit a tight red sequence of galaxies \citep[e.g.,][]{Zirm2008,Tanaka2010}. Here, we plot the color-magnitude diagram for members in QO-1000. Figure \ref{fig:color-mag} shows $J-K_s$ vs. $K_s$ from the COSMOS2020 catalog. $J-K_s$ is from $2\arcsec$ aperture magnitude, whereas $K_s$ magnitude is the Kron magnitude. We see that quiescent members of QO-1000 form a red sequence at $J-K_s\sim1.85$. The spectroscopic quiescent members have a tighter correlation, whereas photometric quiescent members follow them but with a larger scatter. In contrast, star-forming galaxies at $2.74<z_{\rm phot}<2.94$ within $2\sigma$ contour of quiescent galaxies in Figure \ref{fig:1} (denoted as ``star-forming members") have bluer $J-K_s$ colors and fore/background galaxies at $2.0<z_{\rm phot}<3.5$, too. Such a sequence seen in quiescent members of QO-1000 does not exist if we randomly select galaxies at $2<z_{\rm phot}<3.5$ from the entire COSMOS field except QO-1000.
\par We constrain the formation redshift of quiescent members with the model of \citet{Bruzual2003}. The model assumes the single burst star formation with varying formation redshift. The solar metallicity is assumed, supported by the literature on quiescent galaxies at $z>1.5$ \citep[][]{Onodera2015, Saracco2020}. Fitting them to the spectroscopic quiescent members gives their formation redshift close to $z_{\rm form} = 3.7$. The results do not change if we also include photometric quiescent members.
\par This is the first time to see the red sequence in overdense regions with spectroscopically confirmed galaxies at $z\sim3$, whereas some studies had explored with photometrically selected galaxies \citep[e.g.,][]{Kodama2007, Uchimoto2012, Koyama2013}. This analysis also supports that QO-1000 is a concentrated structure of quiescent galaxies. We note that some bright ($K_s<22.5$) galaxies have similar red color ($1.5<J-K_s<2$) but are not classified as quiescent members. They have photometric redshift slightly offset from the targeted redshift slice ($2.74<z_{\rm phot}<2.94$), but may be located in QO-1000 due to its uncertainty.
\subsection{Halo mass estimation} \label{sec:3-4}
\par We attempt to estimate the host halo mass $M_{\rm halo}$ of QO-1000 with two methods referring to similar studies \citep[e.g.,][]{Daddi2021,Sillassen2022}. (1) We convert the stellar mass of the most massive galaxy to its host halo mass. Based on the stellar-to-halo mass ratio in \citet{Behroozi2013}, we derive it as $\log{(M_{\rm halo}/M_\odot)}>13.5$. If we use the recent observationally determined value in \citet{Shuntov2022}, it is estimated as $\log{(M_{\rm halo}/M_\odot)}>13.2$, considering the uncertainty of the stellar-to-halo mass ratio. (2) Assuming the relationship between the total stellar mass and the halo mass for $z\sim1$ clusters \citep{vanderBurg2014}, we convert the sum of the stellar mass of spectroscopically confirmed quiescent members to the host halo mass. Since the distance between QG280611 and others is much larger than the virial radius of the cluster-like halo mass ($\log{(M_{\rm halo}/M_\odot)}\sim14$), QG280611 is not included here, and we assume that the other three are located in the same halo. In this method, the halo mass is estimated as $\log{(M_{\rm halo}/M_\odot)}=13.6$. We note that other galaxies might be in this halo. Therefore, this could be the lower limit.
\par Employing the loosest lower limit, we argue that the halo mass of QO-1000 is $\log{(M_{\rm halo}/M_\odot)}>13.2$.
\subsection{Comparison with IllustrisTNG simulation} \label{sec:3-5}
\par We explore whether there is any structure similar to QO-1000 in the IllustrisTNG simulation \citep[][]{Nelson2019} and its evolution in terms of halo mass.
\par We use the TNG-300 ($205/h$ cMpc box), the largest simulation box. We focus on Snapshot 27 ($z=2.73$), which has the closest redshift to QO-1000. From its subhalo catalog, quiescent galaxies are selected by imposing $\log{\rm (sSFR/yr^{-1})}<-9.5$, which is derived within twice the stellar half-mass radius. To imitate the observation, SFR is averaged over 10Myr \citep[c.f.,][]{Valentino2020}. In addition, the stellar mass threshold of $\log{(M_\star/M_\odot)}>11.0$ is also imposed since all of our spectroscopic confirmed quiescent members satisfy it.
\par From this quiescent galaxy sample, we search structures like QO-1000. The criteria are that (1) three quiescent galaxies should be within $1\ {\rm pMpc}$ of each other, which corresponds to the distance between three galaxies in the central region of QO-1000 (QG306415, QG281561, QG301560), and that (2) in addition to these galaxies, there should be at least one additional quiescent galaxy within $8\ {\rm pMpc}$ from them, which corresponds to the distance from QG280611 to the central galaxies. As a result, we find two structures satisfying the above criteria. There are no systems with more than three QGs in the center (within $1\ {\rm pMpc}$). {\rm Thus, o}ur QO-1000 is the most overdense structure of quiescent galaxies which TNG-300 can reproduce. Note that quiescent galaxies there at Snapshot 27 keep a low sSFR at lower redshift. 
\par
\begin{figure}
    \centering
    \includegraphics[width=9cm]{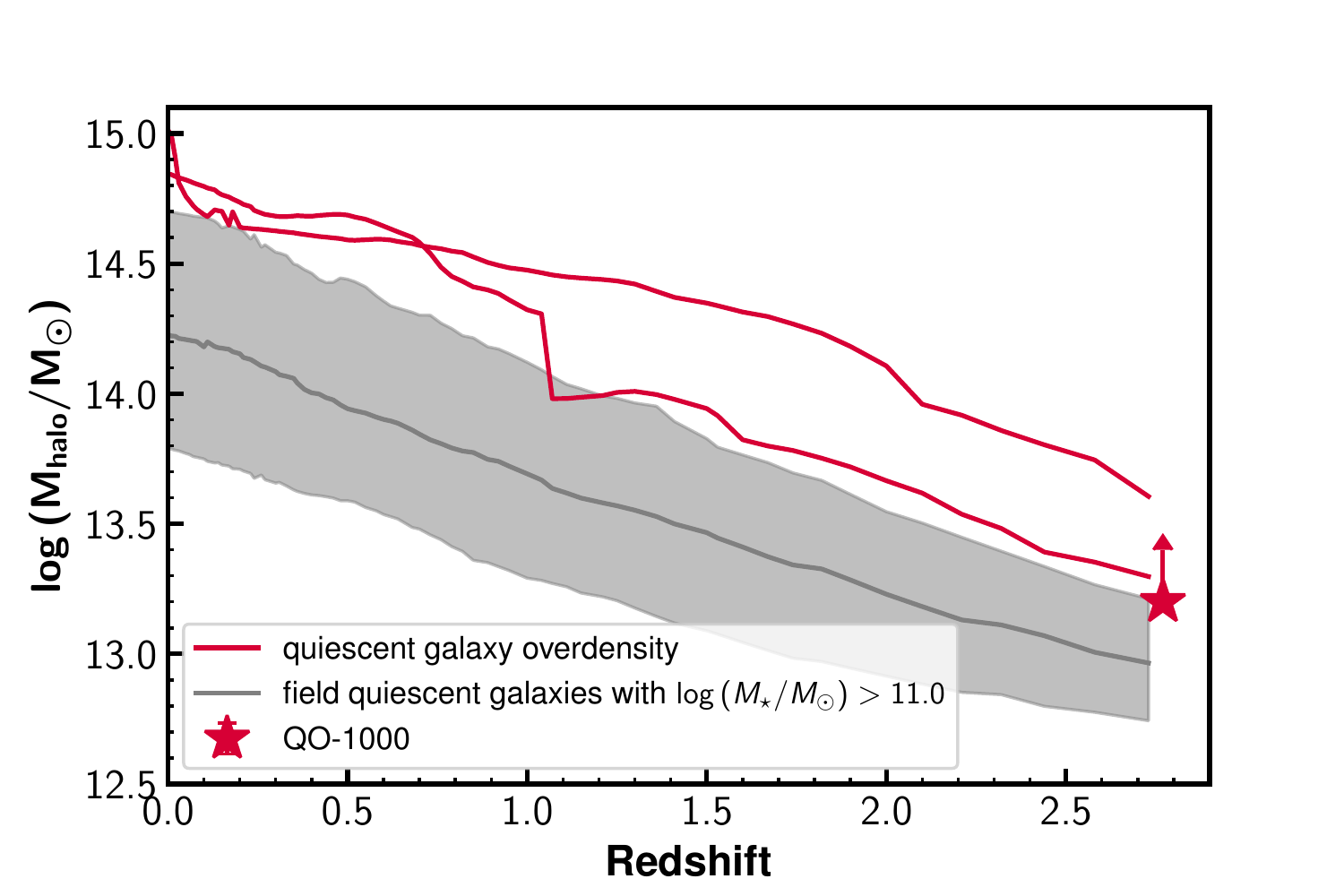}
    \caption{Host halo mass evolution of the most massive quiescent galaxies in two structures similar to QO-1000 in TNG-300 (red lines). The median halo mass evolution of field quiescent galaxies with the same stellar mass in TNG-300 is shown as the gray line. The gray-hatched region corresponds to the range between their 16th and 84th percentiles. The lower limit of the halo mass of QO-1000 (Section \ref{sec:3-4}) is shown in the red star.}
    \label{fig:3}
\end{figure}

 The evolution of their host halo mass is derived by tracing the merger tree. Some quiescent galaxies in the above-selected structures do not reside in the same halo at Snapshot 27. Therefore, we focus on the evolution of the host halo of the most massive galaxies of the structures at Snapshot 27. All quiescent galaxies in the center of the above-selected structures reside in a single halo at $z=0$. As the proxy for the dark matter halo mass $M_{\rm halo}$, we use the mass within the radius where the halo has an overdensity equal to the spherical collapse model threshold defined in \citet{Bryan1998} ({\tt Group\_M\_TopHat200}). Figure \ref{fig:3} shows the evolution of their $M_{\rm halo}$ and those of the other quiescent galaxies with the same stellar mass. Halos of the selected two overdense structures at $z=2.73$ are more massive than the 84th percentile of those of individual quiescent galaxies with similar stellar mass. They are expected to grow into $\log{(M_{\rm halo}/M_\odot)}=14.8,\ 15.0$ at $z=0$. This halo mass evolution is consistent with the lower limit of the halo mass of QO-1000 at $z=2.77$. These mean that the IllustrisTNG simulation supports the existence of an overdense structure of quiescent galaxies at $z\sim3$ and suggests that this QO-1000 will evolve to a Coma-like massive cluster by $z=0$. In other words, QO-1000 is likely to be a protocluster of quiescent galaxies.
\subsection{From star-forming protoclusters to quenched clusters}
\par The high number density of the spectroscopic confirmed quiescent galaxies (Section \ref{sec:3-2}) and the comparison with IllustrisTNG (Section \ref{sec:3-5}) implies that QO-1000 is likely to be a protocluster consisting of quiescent galaxies. This protocluster has the largest number of spectroscopic confirmed quiescent galaxies at $z\sim3$ reported so far. Moreover, this structure has the red sequence of member galaxies and a higher quiescent fraction than the average at the same redshift. These points imply that this structure is more mature than the most known protoclusters, and many of its members are already in the quiescent phase even at $z\sim3$.
\par The high stellar mass of the quiescent galaxies in QO-1000, especially spectroscopically confirmed ones ($\log{(M_\star/M_\odot)}>11.0$), implies that they should have experienced intense star formation, like submillimeter galaxies, and quenched rapidly. Recently, protoclusters of dusty star-forming galaxies have been found at $z\sim4$ \citep[e.g.,][]{Ivison2016, Miller2018}. Many of their member galaxies are as massive as $\log{(M_\star/M_\odot)}\sim11$ \citep[e.g.,][]{Long2020, Rotermund2021}. Thus, such an overdense structure of bursty star-forming galaxies might be the progenitor of QO-1000. Assuming the monotonic decrease of star formation of (proto)cluster galaxies and based on the constraint on the halo mass (Section \ref{sec:3-4}), QO-1000 is likely to be a transition phase from star-forming protoclusters to local quiescent clusters. In this respect, the concentrated region of three spectroscopically confirmed quiescent galaxies might be a protocluster core and will evolve into the center of a cluster. 
\section{Summary} \label{sec:4}
\par In this letter, we report the discovery of an overdense structure of massive quiescent galaxies at $z=2.77$, QO-1000. The $4.2\sigma$ overdensity first finds it in the overdensity map of quiescent galaxies in the COSMOS field. The Keck/MOSFIRE $H$ band spectroscopy detects multiple Balmer absorption lines in the spectra of four galaxies, which implies their redshift is $z=2.76-2.79$. The existence of four massive ($\log{(M_\star/M_\odot)}>11.0$) galaxies with significantly low sSFR ($-10.6<\log{\rm (sSFR/yr^{-1})}<-9.7$) suggests that this structure is at least 68 times denser in quiescent galaxies than in the general field. Thus, we spectroscopically confirm that this structure is a dense structure of quiescent galaxies at $z=2.77$. The quiescent fraction is $\sim3$ times higher than the average value at this redshift. Moreover, there is a red sequence of quiescent galaxies. These also support that quiescent galaxies dominate this structure.
\par The high stellar mass of spectroscopically confirmed quiescent galaxies implies they are hosted in a massive halo with $\log{(M_{\rm halo}/M_\odot)}>13.2$. Compared with the IllustrisTNG simulation, this structure is likely to be hosted by a much more massive halo than the other typical quiescent galaxies with the same stellar mass and going to evolve into a cluster with $\log{(M_{\rm halo}/M_\odot)} \geq14.8$ by $z=0$. Our results imply that this structure is a more mature protocluster than most known protoclusters and is likely in a transition phase from star-forming protoclusters to local quenched clusters. 
\par This discovery confirms that even at $z\sim3$, protocluster galaxies can be quenched, and quiescent galaxies can form an overdense structure. This structure will be an ideal laboratory to explore the evolutionary history of (proto)clusters and galaxies therein. More detailed investigations of member quiescent galaxies in this structure will be conducted in our future studies, such as constraining star formation history based on the spectra and multi-band photometry, investigating morphology using HST/F160W images \citep[3D-DASH][]{Mowla2022}, estimating dynamical mass, and comparing them with simulations. 
\begin{acknowledgments}
\par We appreciate the anonymous referee for helpful comments and suggestions that improved the manuscript. K.I acknowledges Dr. Shuowen Jin, Dr. Kazuhiro Shimasaku, and Mr. Makoto Ando for the discussion about this letter. This work was supported by JSPS KAKENHI Grant Numbers JP 22J00495. O.I. acknowledges the funding of the French Agence Nationale de la Recherche for the project iMAGE (grant ANR-22-CE31-0007). The Cosmic Dawn Center (DAWN) is funded by the Danish National Research Foundation under grant No. 140. The french COSMOS team is supported by the Centre National d'Etudes Spatiales (CNES). The data presented herein were obtained at the W. M. Keck Observatory, which is operated as a scientific partnership among the California Institute of Technology, the University of California and the National Aeronautics and Space Administration. The Observatory was made possible by the generous financial support of the W. M. Keck Foundation. The authors wish to recognize and acknowledge the very significant cultural role and reverence that the summit of Maunakea has always had within the indigenous Hawaiian community. We are most fortunate to have the opportunity to conduct observations from this mountain.
\end{acknowledgments}
\vspace{5mm}
\facilities{Keck:I (MOSFIRE)}
\software{Astropy \citep{Robitaille2013,Price-Whelan2018}, Matplotlib \citep{Hunter2007}, MIZUKI \citep{Tanaka2015}, MOSFIRE DRP, numba \citep{Lam2015}, numpy \citep{Harris2020}, pandas \citep{Mckinney2010}, SLINEFIT \citep{Schreiber2018}}

\bibliographystyle{aasjournal}

\end{document}